\documentclass[prl,aps,twocolumn]{revtex4}
\usepackage{epsfig}
\usepackage{bm}
\newcommand{\B}[1]{{\bm{#1}}}
\newcommand{\C}[1]{{\mathcal{#1}}}

\newcommand{\Onecol} {\begin{widetext} \onecolumngrid} %% 2 -> 1
\newcommand{\Twocol} {\end{widetext} \twocolumngrid} %% 1 -> 2

\newcommand{\be}{\begin{equation}}
\newcommand{\ba}{\begin{array}}
\newcommand{\bea}{\begin{eqnarray}}
\newcommand{\bfi}{\begin{figure}}
\newcommand{\ee}{\end{equation}}
\newcommand{\ea}{\end{array}}
\newcommand{\eea}{\end{eqnarray}}
\newcommand{\efi}{\end{figure}}

\begin{document}
\title{Theory of concentration dependence 
in drag reduction by polymers and of the MDR asymptote}
\author{Roberto Benzi$^{1}$, Emily S.C. Ching$^2$, 
Nizan Horesh$^{2,3}$ and Itamar
Procaccia$^{2,3}$}
\affiliation{$^1$ Dip. di Fisica and INFM, Universit\`a ``Tor
Vergata", Via della Ricerca Scientifica 1, I-00133 Roma, Italy\\
$^2$ Dept. of Physics, The Chinese University of Hong Kong,
Shatin, Hong Kong\\ $^3$ Dept. of Chemical Physics, The Weizmann
Institute of Science, Rehovot,  76100 Israel}
%\pacs{47.27-i, 47.27.Nz, 47.27.Ak}
\begin{abstract}
A simple model of the effect of polymer concentration on the amount of drag reduction in turbulence
is presented, simulated and analyzed.  The qualitative phase diagram of drag coefficient
vs. Reynolds number (Re) is recaptured in this model, including the theoretically elusive onset
of drag reduction and the Maximum
Drag Reduction (MDR) asymptote. The Re-dependent drag and the MDR are analytically
explained, and the dependence of the amount of drag on material parameters is rationalized.
\vskip 0.2cm
\end{abstract}
\maketitle
%%%%%%%%%%%%%%%%%%%%%%%%%%%%%%%%%%%%%%%%%%%%%%%%%%%%%%%%%%%%%%%%%%%%%%%%%%%%
``Drag reduction" refers to the intriguing phenomenon when the
addition of few tens of parts per million (by weight) of
long-chain polymers to turbulent fluids can bring about a
reduction of the friction drag by up to 80\%
\cite{75Vir,97VSW,00SW}. The phenomenon is well documented since
Toms discovered it accidentally in 1946 while studying the
degradation of polymers. The pioneering work of Virk
\cite{75Vir,97VSW} had systematized and organized a huge amount
of experimental information, but the fundamtental mechanism for
the phenomenon has remained under debate for a long time
\cite{69Lu,90Ge,00SW}. All the {\em experimental} and many of the
{\em numerical} \cite{97THKN,98DSB,00ACP,02ACLPP} investigations
of drag reduction focused on channel and pipe geometries;
recently however it had been discovered by {\em numerical
simulations} \cite{03ACBP} of model equations of viscoelastic
flows (like the FENE-P model) that drag reduction appears also in
homogeneous and isotropic turbulence when seeded with polymers.
This brought about a new focus to the search for the mechanism
for drag reduction, since the analysis of model equations without
wall effects should suffice to uncover a mechanism. Indeed, in a
recent paper \cite{03BDGP} the FENE-P equations were simplified
further to a shell model of viscoelastic flow which was shown to
exhibit drag reduction whose mechanism could be fully explored
analytically. In this Letter we present additional crucial
progress where we demonstrate and explain two of the most
prominent (and least understood) characteristics of drag
reduction, i.e. the onset [as a function of Reynolds number (Re)]
and the Maximal Drag Reduction (MDR) asymptote.
%%%%%%%%%%%%%%%%%%%%%%%%%%%%%%%%%%%%%%%%%%%%%%%%%%%
\begin{figure}
\centering
\includegraphics[width=0.50\textwidth]{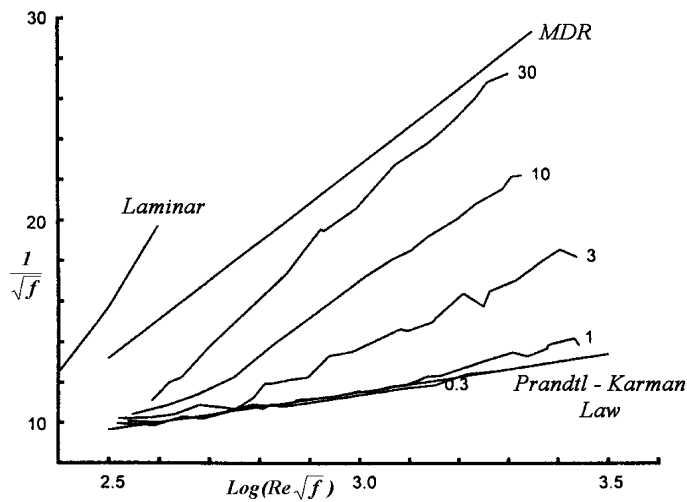}
\caption{Drag reduction in Prandtl-Karman coordinates
\cite{97VSW}. As a function of Re the drag exhibits a
(concentration independent) transition to drag reduction. The
amount of drag reduction depends on the concentration until the
asymptote denoted by MDR is reached. The Prandtl-Karman law is
the Re-dependent drag of the neat fluid. The numbers indicate
concentrations of the polymer additive in wppm.}
\label{cartoonnear}
\end{figure}
%%%%%%%%%%%%%%%%%%%%%%%%%%%%%%%%%%%%%%%%%%%%%%%%%

To set up the issues we reproduce in Fig.1 a typical experimental
figure from  Ref. \cite{97VSW} which refers to the dependence of
the friction (or drag) coefficient in pipe flows on Re. For a pipe
of radius $R$ and length $L$, with $\Delta p$, $\rho$ and $U$
being the pressure drop across $L$, the fluid density and the mean
velocity over a section, the drag coefficient $f$ reads
\begin{equation}
f=\frac{\Delta p}{\rho U^2}\frac{R}{L}\quad \text{(Pipe flow)}\ . \label{deff}
\end{equation}
``Drag reduction" is tantamount to, say, an increase in the
throughput $U$ for a fixed pressure drop $\Delta p$ when polymer
is added to the working fluid. In Fig. 1 one sees that for low Re
there is no drag reduction: the drag coefficient of pure water is
unchanged by the addition of small concentration of polymers.
Then there is a sharp onset of drag reduction at a value of Re
that does not depend on the concentration. From this point on the
amount of drag reduction depends both on the concentration of the
polymer and on Re. It was shown by Virk however that the amount
of drag reduction asymptotes to an apparently universal curve that
cannot be exceeded by increasing the concentration further. This
asymptote is referred to as the MDR, and was claimed to be
insensitive to the nature of the polymer used in the experiments.
In spite of the ample experimental evidence, both the onset and
the existence of the MDR have not been theoretically understood.
In this Letter we wish to close this gap.

Our strategy is to explore simulationally and analytically
simplified models of viscoelastic flows which in spite of the
simplification still represent the robust properties that we are
after. As is well known, viscoelastic flows are represented well
by hydrodynamic equations in which the effect of the polymer
enters in the form of a ``conformation tensor" $\B R(\B r,t)$
which stems from the ensemble average of the diadic product of
the end-to-end distance of the polymer chains. Flexibility and
finite extendability of the polymer chains are reflected by the
relaxation time $\tau$ and the Peterlin function $P(\B r,t)$
which appear in the equation of motion for $\B R$:
\begin{eqnarray}
\frac{\partial R_{\alpha\beta}}{\partial t}+(\B u\cdot \B \nabla)
R_{\alpha\beta}
&&=\frac{\partial u_\alpha}{\partial r_\gamma}R_{\gamma\beta}
+R_{\alpha\gamma}\frac{\partial u_\beta}{\partial r_\gamma}\nonumber\\
&&-\frac{1}{\tau}\left[ P(\B r,t) R_{\alpha\beta} -\rho_0^2
\delta_{\alpha\beta} \right]\label{EqR}\\
P(\B r,t)&&=(\rho_m^2-\rho_0^2)/(\rho_m^2 -R_{\gamma\gamma})
\end{eqnarray}
In these equations $\rho^2_m$ and $\rho^2_0$ refer to the maximal
and the equilibrium values of the trace $R_{\gamma\gamma}$. Since
in most applications $\rho_m\gg \rho_0$ the Peterlin function can
be also written approximately as $P(\B r,t)\approx 1/(1 -\alpha
R_{\gamma\gamma})$ where $\alpha=\rho_m^{-2}$. In its turn the
conformation tensor appears in the equations for fluid velocity
$\B u(\B r,t)$ as an additional stress tensor:
\begin{eqnarray}
&&\frac{\partial \B u}{\partial t}+(\B u\cdot \B \nabla) \B u=-\B \nabla p
+\nu_s \nabla^2 \B u +\B \nabla \cdot \B {\C T}+\B F\ , \label{Equ}\\
&&\B {\C T}(\B r,t) = \frac{\nu_p}{\tau}\left[\frac{P(\B r,t)}{\rho_0^2} \B
R(\B r,t) -\B 1 \right] \ .
\end{eqnarray}
Here $\nu_s$ is the viscosity of the neat fluid, $\B F$ is the
forcing and $\nu_p$ is a viscosity parameter which is related to
the concentration of the polymer, i.e. $\nu_p/\nu_s\sim c$ where
$c$ is the volume fraction of the polymer. Note that the tensor
field can be rescaled to get rid of the parameter $\alpha$ in the
Peterlin function, $\tilde R_{\alpha\beta}=\alpha
R_{\alpha\beta}$ with the only consequence of rescaling the
parameter $\rho_0$ accordingly. These equations were simulated on
the computer in a channel or pipe geometry, reproducing
faithfully the characteristics of drag reduction in experiments
\cite{97THKN,98DSB,00ACP}. It should be pointed out however that
even for present day computers simulating these equations is
quite tasking. We therefore simplify the model further.

In developing a simple model we are led by the following ideas.
First, it should be pointed out that all the nonlinear terms
involving the tensor field $\B R(\B r,t)$ can be reproduced by
writing an equation of motion for a vector field $\B B(\B r,t)$,
and interpreting $R_{\alpha\beta}$ as the diadic product
$B_\alpha B_\beta$.  The relaxation terms with the Peterlin
function are not automatically reproduced this way, and we need
to add them by hand. Second, we should keep in mind that the
above equations exhibit a generalized energy which is the sum of
the fluid kinetic energy and the polymer free energy. Led by
these consideration we write the following shell model
\cite{03BDGP}, which we refer to as the SabraP model:
\begin{eqnarray}
\frac{d u_n}{d  t} &=& \frac{i}{3} \Phi_n (u,u) -  \frac{i}{3}
\frac{\nu_p}{\tau} P(B)
\Phi_n (B,B) - \nu_s k^{2}_n u_n + F_n , \nonumber\\
\frac{d B_n}{d  t} &=& \frac{i}{3} \Phi_n (u,B) -  \frac{i}{3}
\Phi_n (B,u) - {1 \over \tau} P(B) B_n - \nu_B k_n^2 B_n,\nonumber\\
P(B) &=& {1 \over 1 - \sum_n B_n^* B_{n} } \ . \label{SP}
\end{eqnarray}
In these equations $u_n$ and $B_n$ stand for the Fourier amplitudes $u(k_n)$ and $B(k_n)$ of the two
respective vector fields, but as usual in shell model we take $n=0,1,2,\dots$ and the
wavevectors are limited to the set $k_n=2^n$. The nonlinear interaction terms take
the explicit form
\begin{eqnarray}
&&\Phi_n(u,B) = k_n\Big[(1-b) u_{n+2} B^*_{n+1} +(2+b) u^*_{n+1}
B_{n+2}\Big] \nonumber\\&&+ k_{n-1}\Big[(2b+1) u^*_{n-1}B_{n+1}-(1-b)u_{n+1}B^*_{n-1} \Big]\nonumber\\
&&+k_{n-2}\Big[(2+b)u_{n-1}B_{n-2}+(2b+1)u_{n-2}B_{n-1}\Big]\ ,
\end{eqnarray}
with an obvious simplification for $\Phi_n(u,u)$ and
$\Phi_n(B,B)$. Here $b$ is a parameter taken below to be $-0.2$.
In accordance with the generalized energy of the FENE-P model,
also our shell model has the total energy
\begin{equation}
E \equiv {1 \over 2} \sum_n |u_n|^2 - {1 \over 2} {\nu_p \over \tau}
\ln \left(1-\sum_n |B_n|^2\right)\ .
\end{equation}
The second term in the generalized energy contributes to the
dissipation a positive definite term of the form
$(\nu_p/\tau^2)P^2(B) \sum_n|B_n|^2$. With $\nu_p=0$ the first of
Eqs. \ref{SP} reduces to the well-studied Sabra model of
Newtonian turbulence \cite{98LPPPV}. As in the FENE-P case we
consider $\nu_p/\nu_s$ to be $c$. All the simulations below are
performed with a constant rate of energy input, choosing
$F_n=\phi/u^*_n$ for $n=0,1$ and zero otherwise.
%%%%%%%%%%%%%%%%%%%%%%%%%%%%%%%%%%%%%%
\begin{figure}
\centering
\includegraphics[width=.5\textwidth]{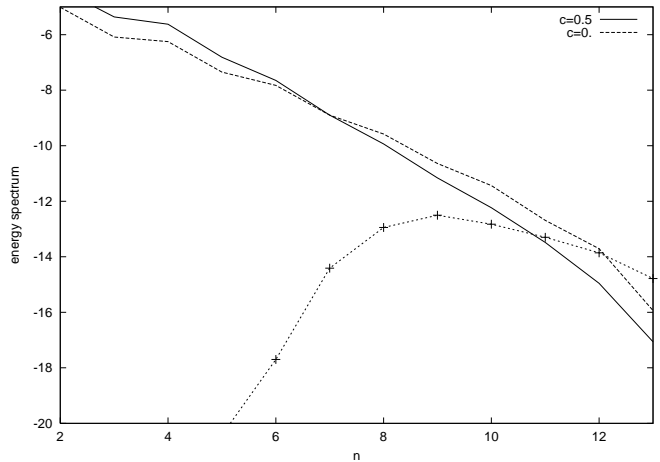}
\caption{Power spectra of the SabraP model (line) and the Sabra
model (dashed line) for $\phi=0.001$, $\nu_s=10^{-6}$ and
$\tau=0.4$. The dashed line with symbols represents the power
spectrum of the $B_n$ field.} \label{logspectrum}
\end{figure}
%%%%%%%%%%%%%%%%%%%%%%%%%%%%%%%%%%%%%%%%%%%%%%%%%

In \cite{03BDGP} it was shown that this shell model exhibits drag
reduction, and the mechanism for the phenomenon was elucidated.
The basic phenomenon is exhibited well by the spectra of the
$u_n$ and $B_n$ fields which are presented at one value of the
parameters in Fig. \ref{logspectrum}. The spectra for the Sabra
model (dashed line) and the SabraP model (line) are compared for
the same amount of power input per unit time. The discussion
\cite{03BDGP} of the spectra revolves around the typical Lumley
scale $k_c$ which is determined by the condition \cite{69Lu}
\begin{equation}
u(k_c) k_c\approx \tau^{-1} \ . \label{defkc}
\end{equation}
For $k_n\gg k_c$ the decay time $\tau$ becomes irrelevant for the
dynamics of $B_n$. The nonlinear interaction between $u_n$ and
$B_n$ at these scales results in both of them having the same
spectral exponent which is also the same as that of the Sabra
model. The amplitude of the $u_n$ spectrum is however smaller in
the SabraP model compared to the Sabra case, since the $B_n$
field adds to the dissipation. On the other hand, for $k_n\ll
k_c$ the $B_n$ field is exponentially suppressed by its decay due
to $\tau$, and the spectral exponent of $u_n$ is again as in the
Sabra model. Drag reduction comes about due to the interactions at
length scales of the order of $k_c$ which force a strong tilt in
the $u_n$ spectrum there, causing it to cross the Sabra spectrum,
leading to an increase in the amplitude of the energy containing
scale. This is why the kinetic energy is increasing for the same
amount of power input, and hence drag reduction. Note that a very
similar spectral cross-over had been documented also for the
FENE-P model in channel flow simulations \cite{02ACLPP}

\begin{figure}
\centering
\includegraphics[width=.5\textwidth]{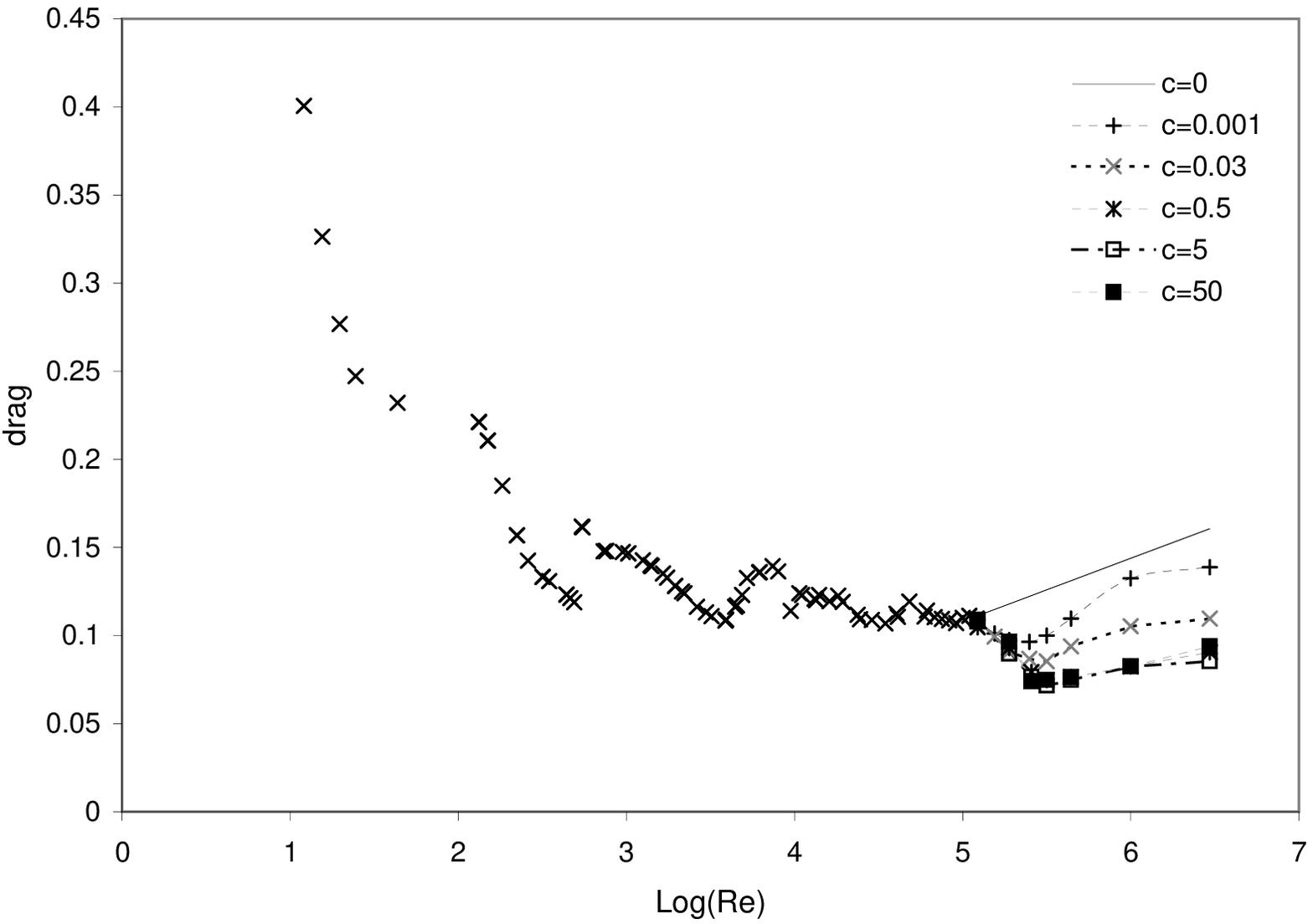}
\includegraphics[width=.5\textwidth]{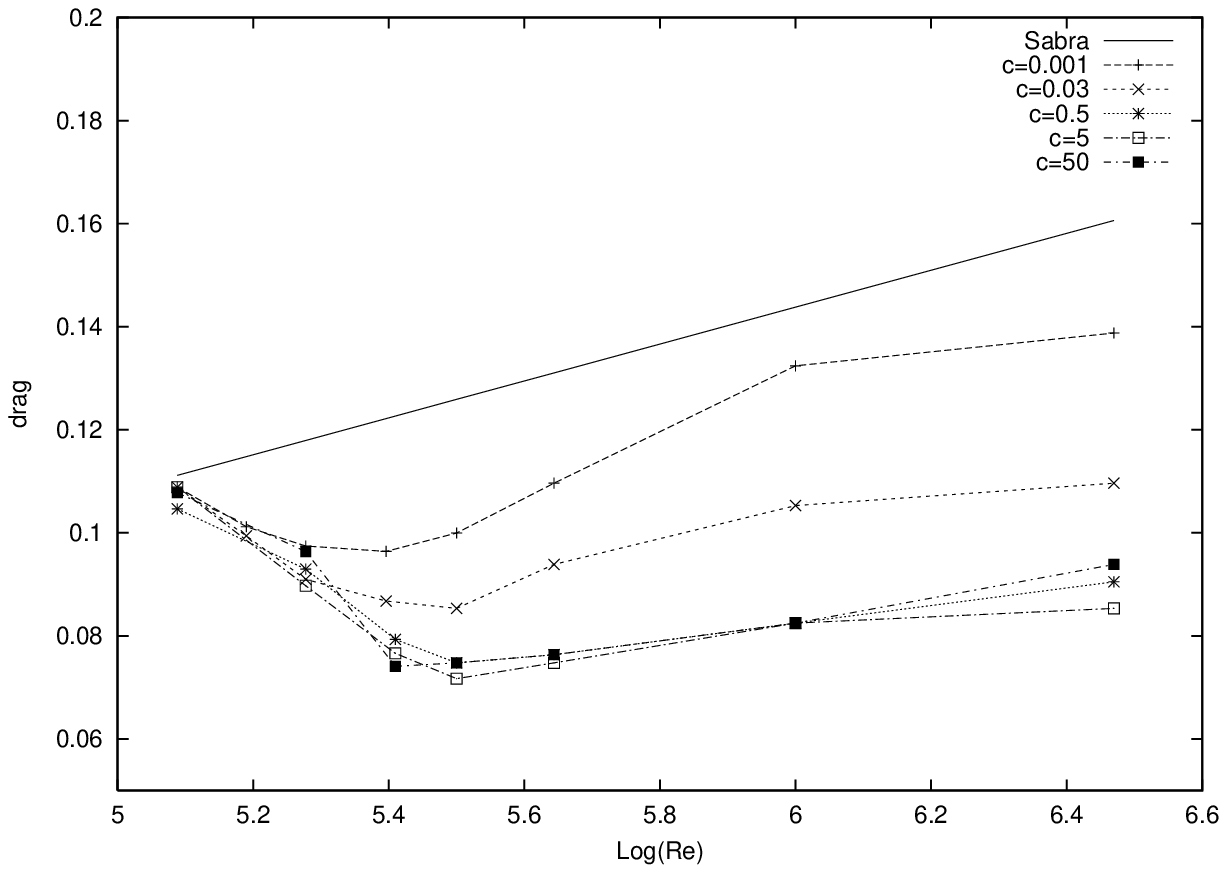}
\caption{Upper panel: Drag as a function of $\log_{10}$(Re)
including the laminar and the turbulent regimes. Both regimes
agree with the Eqs.(\ref{fsmallRe})-(\ref{flargeRe}). Lower panel:
Blow up of the turbulent regime. In both panels the upper
straight line indicates the drag of the neat fluid, whereas the
MDR is seen as the convergence of the drag data for large
concentrations. } \label{c-Re}
\end{figure}

The qualitative phenomena that we are about to explain in this Letter are
demonstrated in the simulational results presented in Fig. \ref{c-Re}.
%%%%%%%%%%%%%%%%%%%%%%%%%%%%%%%%%%%%%%
%%%%%%%%%%%%%%%%%%%%%%%%%%%%%%%%%%%%%%%%%%%%%%%%%
Here we show the drag coefficient $f$ as a function of Re for the
Sabra and for the SabraP models for various values of the
concentration. The drag coefficient is computed in analogy to Eq.
(\ref{deff}) as
\begin{equation}
f\equiv \frac{\sum_n F_n u^*_n}{\Big(\sum_n |u_n|^2\Big)^{3/2}k_0} \ . \quad \text{(Our model)}
\end{equation}
We observe all the phenomena discovered by Virk: (i) For the
model of the neat fluid the drag has a laminar branch and a
turbulent branch, with a sharp transition between them. (ii) For
the model of the viscoelastic flow in the laminar region there is
no drag reduction; the laminar branch is not changed by the
addition of polymer with any concentration. (iii) Drag reduction
has an onset that is independent of the concentration of the
polymer. (iv) As the concentration increases the amount of drag
reduction increases, but (v) there exists an asymptote which is
not exceeded when the concentration is increased. In other words,
our simple model appears to reproduce extremely well the
phenomena that were uncovered in so many experiments as
summarized by Virk.

Next we explain all these observations. First we rationalize the
Re-dependence of the friction factor in the Sabra model of the
neat fluid. For low Re the nonlinear terms $\Phi_n(u,u)$ are
negligible compared to the viscous term. Forcing only on the
largest scale $k_0$ we can evaluate,
\begin{equation}
\nu_s k_0^2 u_0 \approx F_0 \quad \rightarrow~f\sim
\frac{\nu_sk_0}{|u_0|}= {\rm Re}^{-1} \quad \text{Re small} \  .
\label{fsmallRe}
\end{equation}
For large Re we have the exact result \cite{98LPPPV} that the
third order correlation function  $S_n^{(3)}\equiv \Im \langle
u_{n-1} u_n u^*_{n+1}\rangle =C \bar \epsilon /k_n$ with $\bar
\epsilon$ being the mean energy flux and $C$ a known constant
(the analog of the 4/5th law for Navier-Stokes turbulence). We
therefore expect the friction factor to tend to a constant value
for large Re (up to terms $\sim \log$Re),
\begin{equation}
f\sim Re^{0} \quad \text{Re large} \  . \label{flargeRe}
\end{equation}
Eqs. (\ref{fsmallRe}) and (\ref{flargeRe}) (which are the analogs
of the Prandtl-Karman law for pipe flows) are well borne out by
the data in Fig. \ref{c-Re} for the model of the neat fluid. The
laminar branch, which is exponential in these coordinates ($f \sim
\exp\{-\log({\rm Re})\})$, is unaffected by changing the
concentration $c$. The transition between the two branches is
expected when turbulence sets in, i.e. for Re such that the
dissipative terms just begin to be overwhelmed by the nonlinear
interactions. Thus point (i) is understood. Note that similar
arguments will hold for the FENE-P equations in homogeneous
flows. Points (ii) and (iii) are explained as follows; we said
above that drag reduction comes about due to the interaction
between the two dynamical fields at scale of the order of $k_c$.
Clearly, as long as $k_c$ exceeds the dissipative scale $k_d$ of
the the $u_n$ field, no interaction between the two field can be
of any significance. Since $k_d$ is of the order of $k_d\sim k_0
Re^{3/4}$, we can expect a concentration independent onset of
drag reduction when $k_c\approx k_d$. Using $u_n\sim
u_0(k_n/k_0)^{-1/3}$, $k_c$ can be estimated as $k_c\sim k_0
(\tau u_0k_0)^{-3/2}$, and we end up with a prediction for the
onset of drag reduction when
\begin{equation}
Re \approx (\tau u_0k_0)^{-2} \quad \text {Onset of drag reduction} \ .
\end{equation}
This prediction is well borne out by our simulations (due to the
space constraint we do not display simulations at different values
of $\tau$ and $k_0$). Again we point out that similar arguments
can be presented for the FENE-P model as well.

Point (iv) is obvious - when the concentration increases the
mechanism discovered in \cite{03BDGP} comes into play. What
remains to explain is the asymptotic MDR. This also follows
directly from the analysis of the equations. Consider Eqs.
(\ref{SP}) for two values of the parameter $\nu_p$,
$\nu_p^{(1)}\ll \nu_p^{(2)}$, with $y^2=\nu_p^{(1)}/\nu_p^{(2)}$.
Rescaling $B_n$ according to $B_n=y\tilde B_n$, we see that the
Peterlin function tends to unity when $y\to 0$,
\begin{equation}
P(\tilde B)=\frac{1}{1- y^2\sum_n|\tilde B_n|^2} \to 1\ , \quad
\text {when}~y\to 0 \ .
\end{equation}
When $P(\tilde B)\approx 1$ the dynamical equation for $\tilde
B_n$ is independent of of $y$ due to its linearity, whereas the
$u_n$ equation remains independent due to the rescaling:
\begin{equation}
\frac{i}{3} \frac{\nu_p^{(2)}}{\tau} P(B) \Phi_n (B,B)\to
\frac{i}{3}\frac{\nu_p^{(1)}}{\tau}  P(\tilde B) \Phi_n(\tilde
B_n, \tilde B_n)\ ,
\end{equation}
Thus increasing the concentration
brings to the dynamical equations to an asymptotic concentration invariant form and therefore
to an asymptotic MDR. For the last time we remark that similar rescalings are also
available in the FENE-P equations, making the points discussed here quite general for
any sensible model of viscoelastic flow.

In summary, we have presented a simple model of drag reduction
for which the observed characteristics can be explained on the
basis of the equations of motion. It remains to go back to
channel and pipe simulations of the FENE-P equations to
demonstrate that the discussion presented above includes the main
phenomena observed also there.
%%%%%%%%%%%%%%%%%%%%%%%%%%%%%%%%%
\acknowledgments \vskip 0.5 cm This work was supported in part by
the European Commission under a TMR grant, the Minerva
Foundation, Munich, Germany, and the Naftali and Anna
Backenroth-Bronicki Fund for Research in Chaos and Complexity.
ESCC acknowledges the Hong Kong Research Grants Council for
support (CUHK 4046/02p).
%%%%%%%%%%%%%%%%%%%%%%%%%%%%%%%%%%

\end{document}